\documentclass{aa} 
\usepackage{graphicx}
\usepackage{txfonts}
\usepackage{natbib}
\begin{document}

   \title{Revisiting the mass- and radius-luminosity relations for FGK main-sequence stars\thanks{This paper is dedicated to the memory of Johannes Andersen.}}


   \author{Jo\~ao Fernandes
          \inst{1},\inst{2}
          \and
          Ricardo Gafeira\inst{3}
 \and Johannes Andersen\inst{4}\thanks{Johannes Andersen passed away on 28 April 2020.}
          }

   \institute{Univ Coimbra, CITEUC, Geophysical and Astronomical Observatory, Coimbra, Portugal\\
              \email{jmfernan@mat.uc.pt}
         \and
             Univ Coimbra, CITEUC, Department of Mathematics, Coimbra, Portugal
\and
Univ Coimbra, CITEUC, Department of Physics, Coimbra, Portugal\\
              \email{gafeira@mat.uc.pt}
 \and 
 Niels Bohr Institute, Blegdamsvej 17, DK-2100, Copenhagen, Denmark\\
 \email{ja@nbi.ku.dk}
             }

   \date{Received ; accepted }


  \abstract
   {Scaling relations are very useful tools for estimating unknown stellar quantities. Within this framework, eclipsing binaries are ideal for this goal because their mass and radius are known with a very good level of accuracy, leading to improved constraints on the models.  }
   {We aim to provide empirical relations for the mass and radius as function of luminosity, metallicity, and age. We investigate, in particular, the impact of metallicity and age on those relations.}
   {We used a multi-dimensional fit approach based on the data from $DEBCat$, an updated catalogue of eclipsing binary observations such as mass, radius, luminosity, effective temperature, gravity, and metallicity. We used the {PARAM web interface for the Bayesian estimation of stellar parameters, along with} the stellar evolutionary code MESA to estimate the binary age, assuming a coeval hypothesis for both members.}
   {We derived the mass and radius-luminosity-metallicity-age relations using 56 stars, {with metallicity and mass in the range $-0.34<$[Fe/H]$<0.27$ and $0.66<M/M_\odot<1.8$}. With that, the observed mass and radius are reproduced with an accuracy of $3.5\%$ and $5.9\%,$ respectively, which is consistent with the other results in literature.}
   {We conclude that including the age in such relations increases the quality of the fit, particularly in terms of the mass, as compared to the radius.   On the other hand, as other authors have noted, we observed an higher dispersion on the mass relation than in that of the radius. We propose that this is due to a stellar age effect.}

   \keywords{stars:binaries: eclipsing; evolution; fundamental parameters; solar-type}

    \titlerunning{Mass/radius-luminosity relations}
    \authorrunning{Fernandes, Gafeira and Andersen}
   \maketitle

%

\section{Introduction}
Empirical stellar relations establish the dependence between fundamental proprieties of stars in a simple way. These relations, frequently, rely on 'obvious' astrophysical arguments. The most well-known case is the mass-luminosity relation (MLR). The MLR is an expected relation given that stellar energy production - and, thus, the luminosity -  is strongly dependent on mass. According to , Jakob Karl Ernst Halm was the first to remark on the existence of the MLR. In the words of G. P. Kuiper, who {points out the relevance of such a relation, remarking on the importance of this relation to progress in the field, it stands as a }'a central problem of theoretical astrophysics' (\citet{1938ApJ....88..472K}. Nowadays, the improvement of the accuracy of mass in stellar binaries, down to 3\% or better, as per, for example, \cite{torres}, allows us to establish precise MLRs (\cite{2018ApJS..237...21M} and references. Another example is the age-metallicity relation, which is based on the simple idea that a gradual chemical enrichment of the interstellar medium over the time is accompanied by an increase of the metallically in stellar generations \cite{2006A&A...453L...9R}. Many more examples can be pointed out: mass-radius relations, MRR \citep{2018MNRAS.479.5491E}, helium-to-metal enrichment \citep{2019MNRAS.483.4678V}, seismic scaling relations \citep{2019MNRAS.486.4612B}, etc. The substantial advantage of these kinds of relations is clearly to enable predictions of unknown observed stellar parameters based on observed ones. 

For instance, in terms of the MLR, for single nearby solar-type stars, the luminosity can be obtained observationally (with a good level of accuracy), but not the mass. A detailed review of stellar-mass determinations can be found in the recent paper by \citet{2020arXiv200610868S}. However, there are strong limitations with regard to the accuracy of those predictions other than the observational errors, that is, the scatter on the relations themselves thanks to different contributions such as evolution, chemical composition, stellar rotation, and activity, etc. Over the last few years, we can identify several works, particularly, on the topic of MLR, which include the contributions of metallicity and the evolutionary effects on the relation. The inclusion of the metallicity and the age lies on the fundamental proprieties of a star as the luminosity and the radius (or the effective temperature) depends not only on mass, but also on age and chemical composition (cf. the so-called Vogt–Russell theorem). It is for this reason that the helium abundance could also be considered. Unfortunately, that is not measurable in the atmospheres of FGK stars and the determinations via theoretical stellar models still carry uncertainties larger than 10\% \citep{2013A&A...553A..62V}. This value can go down to about 5\% if seismic information is included \citep{2021MNRAS.500...54N}. We make note of the findings \cite{2018ApJS..237...21M} as theirs is the most comprehensive analysis on the field to date. Based on 934 stars, they were able to establish 38 revised or new relations of mass or radius as a function of observations as effective temperature ($T_{\rm eff}$), luminosity ($L$), surface gravity ($g$), density ($\rho$), and metallicity ([Fe/H]). Finally, they concluded that the global precision and accuracy obtained is about 10\%, which is similar to previous works with the advantage to provide 'at least one relation for estimating the stellar mass and radius' as a function of different combinations of those observations. They also present an analysis based on machine-learning techniques training a random forest model, concluding that the accuracy increases in relation to standard linear regression. However, none of the presented relations includes the stellar age. The authors explain that such a relation is not useful as the age is not known with good precision or accuracy. 

In an earlier paper, (\citet{2012Ap&SS.341..405G} (hereafter, Paper I), we analysed 26 FGK stars with a known mass, luminosity, metallicity, and age for all members of binary systems, with observational mass errors less than 3\%, taken from \citet{torres}, including the Sun. We derived the MLR taking
into account, as well as the metallicity and age, separately. Our results
show that the inclusion of age and metallicity in the
MLR, for FGK stars, improves the individual mass estimation by 5 \% to 15 \%. The main limitation of that work is the low number of stars. In the present paper, we enlarged our sample to 56 stars and we revisited the above analysis for MLR, extending it also to the radius-luminosity relation (RLR). For that purpose, we used the  {\it{DEBCat}} survey, which is a catalogue of the physical properties of well-studied detached eclipsing binaries \citep{2015ASPC..496..164S}, including $\alpha$ Centauri, the best-known visual binary. We selected the binaries where both components are in the main sequence with FGK spectral types, and mass and radius are known as better as 3\% to 4\%, respectively. 

The goals of the present paper are: (i) to provide relations for mass and radius as a function of luminosity, age, and [Fe/H]; (ii) to quantify the effects of metallicity and age on the MLR and RLR; and (iii) to discuss the accuracy of mass and radius estimation from those relations. The paper is organised as follows. In Sect. 2, we present the observational data for the sample of binaries. We compute the stellar ages and masses for each star by means of theoretical stellar models, assuming both binary members are coeval. In Sect. 3, we derive the MLR and RLR, including the age and metallicity contributions and we discuss the results. In Sect.  4, we summarise our main conclusions.

\section{Eclipsing binaries data and model: Binary age determination}
Our stellar sample is taken from {\it{DEBCat}\footnote{\url{https://www.astro.keele.ac.uk/jkt/debcat/}},} assuming certain criteria that were both observational and theoretical. On the observational side, we chose the stars for which the gravity, effective temperature, luminosity, and metallicity, as well as the respective uncertainties are available for both components. For the visual binary system $\alpha~$Centauri, we took the data from \cite{2017A&A...597A.137K}. On the other hand, we chose the binaries where both components are main-sequence FGK stars and the metallicities are typical for Population I (which means [Fe/H]$>-0.34$ for this sample). When the luminosity class is not explicitly written in {\it{DEBCat}}, we considered a candidate for a main-sequence star if $3.4<\log ~g<4.6$ from \cite{1996BABel.154...13A}. The reason for this choice lies in the stellar models used to compute the mass and age (see below). In spite of the independent age determinations for each binary system (see respective source papers in {\it{DEBCat}}), for the sake of consistency, we chose to recompute the age by means of theoretical stellar evolutionary models. These models are scaled-solar and, thus, $[\alpha/Fe]=0$, which indicates metal-rich stars (\citet{2014A&A...567A...5R}). Moreover, stellar models on that range are based on the input physics to realistically describe  the internal structure and the evolution of the Sun, and thus, they are suitable for reproducing FGK stars, as their observations are well-recovered by means of those theoretical models, such as \citet{lebretonetal08}. Within those criteria, we select 33 binaries, including the visual binary $\alpha ~Cen$. Those binaries are listed in Table~\ref{fgkstars}, except for five of them (see more information below). {The age and the corresponding error for each binary is also presented (error is computed as the average
of the half width of the $68\%$ confidence interval)}. 

\begin{figure}
   \centering
    \includegraphics[width=\hsize]{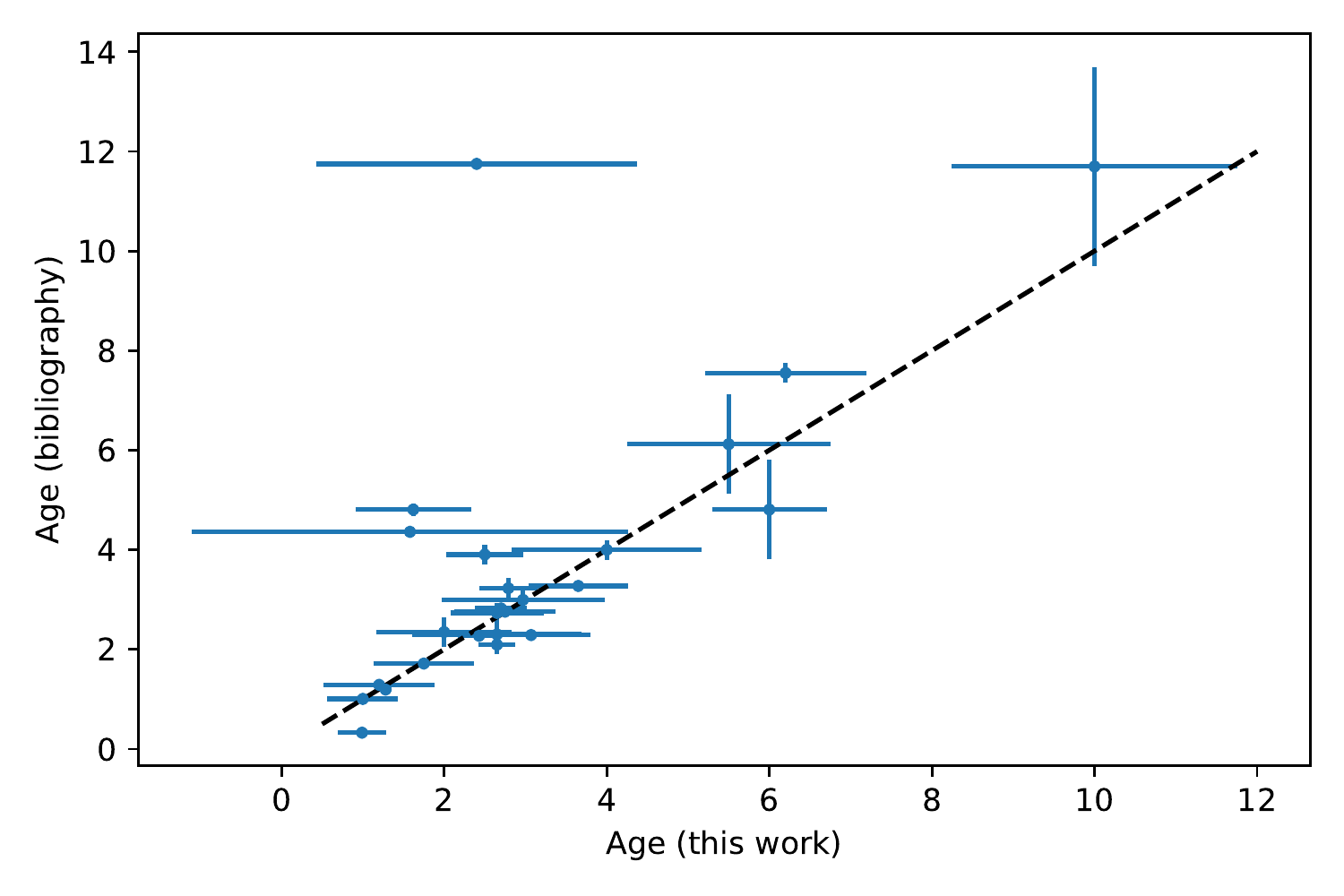}

      \caption{Age of this work versus the referenced age. Dashed line is a one-to-one relation. {The horizontal and vertical blue lines represent respectively the age errors obtained form this work and the observational errors. We did not include any error bars on the cases where the observational error is not determined.}
              }
         \label{graficoage}
   \end{figure}

The stellar age values are presented as function of the corresponding solar value $age_\odot = 4.6~Gyr$ \citep{2015A&A...580A.130B}. In order to compute stellar ages and masses, we used the MESA isochrones from \citet{2017MNRAS.467.1433R} and \cite{2015ApJS..220...15P}, with scaled-solar composition and following the chemical enrichment law, $Y=0.2485+1.007\times Z,$ assuming the solar metal content is $Z_\odot$=0.01756. To achieve this, we used the public input form PARAM 1.4\footnote{\url{http://stev.oapd.inaf.it/cgi-bin/param_1.4}}. Ages were determined using a Bayesian estimation, assuming each star is a single one, taking into account the above observational data for each star ($\{T_{\rm eff},\log~g, L,$[Fe/H]$\}$) {and the respective errors}.  The age of the system is estimated by the average of the age of individual stars. We also present the values for the Sun. Using the same procedure, the solar observations \citep{2016AJ....152...41P} were recovered for a model with $age=4.2904\pm0.0621~Gyr$ and $M/M_\odot=1.0027\pm0.0021$. Following the age and mass determinations, we excluded five binaries for two reasons: there was a 'no solution' from the PARAM platform or the computed solution was 'astrophysically inconsistent'. The label of 'no solution' indicates that the PARAM solver did not converge for the input observational data. This is the case for the following binaries:
\begin{itemize}
  \item ASAS J065134-2211,5: \cite{2019A&A...622A.114H} presents the orbital and physical parameters of this binary. Concerning the analysis by means of stellar models, they use the PARSEC code from \cite{2017ApJ...835...77M} and they fit the primary star in $1\sigma$ for an isochrone of $6.3~Gyr$ in the observational plan $\{(M~vs~T_{\rm eff}),(M~vs~L),(M~vs~R)\}$. However, the authors conclude that for the secondary, the radius is larger and $T_{\rm eff}$ is clearly lower than the models. They claim that this problem could be related to the chromospheric activity of the system.
  \item ASAS J073507-0905,7: \cite{2019A&A...622A.114H} also studied this binary. The analysis was preformed in the same observational plans. Both the proprieties of primary and secondary are only fitted within $3\sigma$ for a very young age (10 to 25 Myr) that could indicate that both stars are still in PMS phase.
  \item NP Per:  \cite{2016AJ....152....2L} deemed it as a PMS eclipsing binary. On the other hand, using the MESA models, the authors claim that the observations of both stars were not fitted with a single isochrone.
\end{itemize}
With regard to the label of a solution that was 'astrophysically inconsistent' indicates that, cumulatively, the ages of both stars are not compatible with a coeval hypothesis (within $1~\sigma$) and the masses of both binary components recovered by the PARAM solver are not compatible with the corresponding observed values, within $1~\sigma$. Using these criteria, two more binaries are excluded:
\begin{itemize}
  \item EF Aqr: Our solutions point to ages for the primary around $2.5\pm0.4~Gyr$ and $14\pm1~Gyr$ for the secondary, which is incompatible with the coeval hypothesis. On the other hand, both observed masses are outside $1\sigma$ of the masses given by the models: A - $1.2440\pm0.0080~vs~1.19\pm0.03$ and B - $0.9460\pm0.0061~vs~0.86\pm0.02$. The difficulties involved in modelling EF Aqr are confirmed by \cite{2012A&A...540A..64V}, as they conclude that different set of stellar evolutionary models, scaled on the Sun, are unable to recover the observations of this binary, mainly $T_{\rm eff}, R, M,$ and $L$. However, an agreement between the models and observations can be obtained if the mixing length parameter of the models is considerably reduced in relation to the solar value ($\alpha_\odot=1.68$ for GRANADA models - \cite{2004A&A...424..919C}): 1.3 for the primary and 1.05 for the secondary. We recall that the models in our paper assume the solar-mixing length parameter. The proposed mixing length parameter reduction can be explained by the stellar high rotational velocity and its impact in the convective energy transport \citep{2009A&A...502..253C}.
  \item V636 Cen: Our solutions propose ages for the primary of around $3.5\pm1~Gyr$ and $14\pm1~Gyr$ for the secondary. So, again, they are incompatible with the coeval hypothesis. On the other hand, both observed masses are outside $1\sigma$ of the masses given by the models: A - $1.051\pm0.0051~vs~0.984\pm0.03$ and B - $0.8540\pm0.0030~vs~0.768\pm0.03$. \cite{2009A&A...502..253C} analysed this binary, drawing a conclusion on the difficulty in modelling the stars using the same kind of arguments as for EF Aqr: the mixing length parameter of both stars must be considerably reduced in order to achieve an agreement between models and observations.
\end{itemize}
In Fig.~\ref{graficoage}, we plot our age determinations for the remaining binaries in comparison to those presented in the corresponding source papers of each binary {and corresponding error bars}. For half of the binaries, the error on age was not published. The mean of the absolute difference {between two samples - the previously published one and ours -} is about $1.3~Gyr$.   \cite{2019MNRAS.482..895S} discuss the age for Gaia benchmark stars. They consider as Rank A stars those 'for which the age range is well-defined and within an interval of a few Gyr'. According to their Table 4, we can infer that for FG dwarfs ranked A, $1.5~Gyr$ error seems realistic and can be compared to the above differences. Assuming this value, all binaries, but one, lies on $1~to~2 \sigma$ of mutual error bars. The outlier is IM Vir. The age determined by \cite{2009ApJ...707..671M} is $2.4~Gyr$ in contrast with our of about $12\pm2~Gyr$. The main reason for that is the fact that \cite{2009ApJ...707..671M} introduced corrections to $T_{\rm eff}$ and to the mixing-length parameter for both stars to mimic the physical effects of the stellar chromospheric activity. Naturally, these corrections were not implemented in our age determinations.

\begin{table*}

\centering\tiny
\begin{tabular}{r r r r r r r r r r r | r r}
\hline\hline
Binary   & star   & $\log M/M_\odot$& error & $\log R/R_\odot$ & error & $\log L/L_\odot$ & error          & $[M/H]$& error & \textbf{Ref.} &$age/age_\odot$ & error \\
\hline
AD~Boo  &       A       &       0.1504  &       0.0028  &       0.2074  &       0.0038  &       0.640   &       0.030   &       0.10    &       0.15 &(1)    &       0.3729  &       0.1351  \\
AD~Boo  &       B       &       0.0824  &       0.0022  &       0.0849  &       0.0036  &       0.280   &       0.040   &               &        &       &               &       \\
AL~Dor  &       A       &       0.0425  &       0.0002  &       0.0496  &       0.0039  &       0.159   &       0.033   &       0.10    &       0.10 &(2)    &       0.6511  &       0.2179  \\
AL~Dor  &       B       &       0.0421  &       0.0002  &       0.0484  &       0.0039  &       0.145   &       0.033   &               &        &       &               &               \\
$\alpha$ Cen    &       A       &       0.0434  &       0.0028  &       0.0878  &       0.0011  &       0.190   &       0.008   &       0.24    &       0.04 &       (3)&    0.9495  &       0.1532  \\
$\alpha$ Cen    &       B       &       -0.0297 &       0.0028  &       -0.0640 &       0.0025  &       -0.303  &       0.021   &               &        &       &               &               \\
BG~Ind  &       A       &       0.1547  &       0.0024  &       0.3598  &       0.0032  &       0.870   &       0.070   &       -0.20   &       0.10& (4)     &       0.5938  &       0.1248  \\
BG~Ind  &       B       &       0.1116  &       0.0027  &       0.2253  &       0.0098  &       0.680   &       0.080   &       & &       &               &               \\
BK~Peg  &       A       &       0.1504  &       0.0021  &       0.2984  &       0.0017  &       0.740   &       0.020   &       -0.12   &       0.07&(5)        &       0.6135  &       0.0694  \\
BK~Peg  &       B       &       0.0993  &       0.0017  &       0.1685  &       0.0050  &       0.490   &       0.030   &               &        &       &               &               \\
BW~Aqr  &       A       &       0.1377  &       0.0057  &       0.2388  &       0.0022  &       0.669   &       0.027   &       -0.07   &       0.11 &(6)    &       0.4939  &       0.0577  \\
BW~Aqr  &       B       &       0.1670  &       0.0053  &       0.3161  &       0.0008  &       0.796   &       0.028   &               &        &       &               &               \\
CO~And  &       A       &       0.1103  &       0.0025  &       0.2373  &       0.0053  &       0.580   &       0.038   &       0.01    &       0.15 &(7)    &       0.7117  &       0.1334  \\
CO~And  &       B       &       0.1019  &       0.0025  &       0.2289  &       0.0044  &       0.572   &       0.038   &               &        &       &               &               \\
EPIC~210822691  &       A       &       0.0374  &       0.0040  &       0.0969  &       0.0035  &       0.245   &       0.028   &       0.00    &       0.05    & (8)&    1.3228  &       0.5111  \\
EPIC~210822691  &       B       &       -0.0177 &       0.0090  &       -0.0269 &       0.0046  &       -0.118  &       0.038   &               &               & &               &               \\
EPIC~219394517  &       A       &       0.0327  &       0.0008  &       0.0233  &       0.0045  &       0.091   &       0.031   &       0.07    &       0.03    & (9)&    0.4999  &       0.2257  \\
EPIC~219394517  &       B       &       0.0278  &       0.0010  &       0.0179  &       0.0050  &       0.066   &       0.031   &               &               & &               &               \\
EW~Ori  &       A       &       0.0693  &       0.0041  &       0.0674  &       0.0019  &       0.220   &       0.030   &       0.05    &       0.09    &       (10)& 0.5105  &       0.1806  \\
EW~Ori  &       B       &       0.0504  &       0.0035  &       0.0402  &       0.0020  &       0.120   &       0.030   &               &               &       &         &               \\
FL~Lyr  &       A       &       0.0829  &       0.0027  &       0.0948  &       0.0080  &       0.330   &       0.095   &       -0.07   &       0.09    &       (11)& 0.9479  &       0.5833  \\
FL~Lyr  &       B       &       -0.0217 &       0.0018  &       -0.0458 &       0.0116  &       -0.180  &       0.080   &               &               &       &         &               \\
GX~Gem  &       A       &       0.1726  &       0.0032  &       0.3668  &       0.0022  &       0.860   &       0.030   &       -0.12   &       0.10    &       (12)& 0.7018  &       0.0788  \\
GX~Gem  &       B       &       0.1664  &       0.0030  &       0.3499  &       0.0023  &       0.820   &       0.030   &               &               &       &         &               \\
IM~Vir  &       A       &       -0.0083 &       0.0053  &       0.0257  &       0.0065  &       -0.012  &       0.034   &       -0.10   &       0.25    &       (13)& 2.5543  &       0.4300  \\
IM~Vir  &       B       &       -0.1776 &       0.0031  &       -0.1669 &       0.0083  &       -0.867  &       0.056   &               &               &       &         &               \\
Kepler-34       &       A       &       0.0203  &       0.0013  &       0.0651  &       0.0011  &       0.173   &       0.033   &       -0.07   &       0.15    &       (14)& 1.3305  &       0.2720  \\
Kepler-34       &       B       &       0.0089  &       0.0009  &       0.0385  &       0.0012  &       0.107   &       0.036   &               &               &       &         &               \\
Kepler-35       &       A       &       -0.0517 &       0.0025  &       0.0122  &       0.0008  &       -0.027  &       0.043   &       -0.34   &       0.20    &       (15)& 2.5429  &       0.3827  \\
Kepler-35       &       B       &       -0.0918 &       0.0024  &       -0.1045 &       0.0012  &       -0.387  &       0.020   &               &               &       &         &               \\
KIC~10031808    &       A       &       0.2408  &       0.0022  &       0.4133  &       0.0034  &       1.185   &       0.027   &       -0.11   &       0.08    &(16) &       0.2593  &       0.0157  \\
KIC~10031808    &       B       &       0.2548  &       0.0031  &       0.4810  &       0.0020  &       1.254   &       0.027   &               &               & &               &               \\
KIC~2557430     &       A       &       0.2279  &       0.0077  &       0.2742  &       0.0046  &       0.899   &       0.038   &       -0.11   &       0.03    &       (17)&0.3651     &       0.2134  \\
KIC~2557430     &       B       &       0.1303  &       0.0064  &       0.0899  &       0.0106  &       0.249   &       0.075   &               &               &       &       &               \\
KIC~3439031     &       A       &       0.0890  &       0.0010  &       0.1471  &       0.0008  &       0.450   &       0.060   &       0.10    &       0.13    &       (18)&0.4972     &       0.1591  \\
KIC~3439031     &       B       &       0.0884  &       0.0009  &       0.1484  &       0.0007  &       0.450   &       0.060   &               &               &       &       &               \\
KIC~7177553     &       A       &       0.0183  &       0.0058  &       -0.0269 &       0.0023  &       -0.055  &       0.040   &       -0.05   &       0.09    &       (19)&0.3702     &       0.2341  \\
KIC~7177553     &       B       &       -0.0061 &       0.0066  &       -0.0264 &       0.0023  &       -0.071  &       0.041   &               &               &       &       &               \\
KIC~7821010     &       A       &       0.1062  &       0.0058  &       0.1059  &       0.0037  &       0.470   &       0.040   &       0.10    &       0.08    &       (20)&0.0712     &       0.0652  \\
KIC~7821010     &       B       &       0.0867  &       0.0057  &       0.0828  &       0.0050  &       0.390   &       0.050   &               &               &       &       &               \\
KX~Cnc  &       A       &       0.0561  &       0.0011  &       0.0269  &       0.0008  &       0.091   &       0.002   &       0.07    &       0.10    &       (21)&0.2186     &       0.0946  \\
KX~Cnc  &       B       &       0.0535  &       0.0012  &       0.0208  &       0.0008  &       0.061   &       0.002   &               &               &       &       &               \\
LL~Aqr  &       A       &       0.0777  &       0.0003  &       0.1209  &       0.0020  &       0.332   &       0.014   &       0.02    &       0.04    &       (22)&0.8486     &       0.1033  \\
LL~Aqr  &       B       &       0.0144  &       0.0003  &       0.0009  &       0.0022  &       -0.019  &       0.016   &               &               &       &       &               \\
LV~Her  &       A       &       0.0766  &       0.0036  &       0.1329  &       0.0038  &       0.349   &       0.044   &       0.08    &       0.21    &       (23)&0.8695     &       0.2541  \\
LV~Her  &       B       &       0.0681  &       0.0030  &       0.1183  &       0.0036  &       0.311   &       0.044   &               &               &       &       &               \\
UX~Men  &       A       &       0.0874  &       0.0005  &       0.1294  &       0.0042  &       0.380   &       0.030   &       0.04    &       0.10    &       (24)&0.5995     &       0.1356  \\
UX~Men  &       B       &       0.0747  &       0.0005  &       0.1052  &       0.0044  &       0.320   &       0.030   &               &               &       &       &               \\
V1130~Tau       &       A       &       0.1159  &       0.0027  &       0.1729  &       0.0029  &       0.590   &       0.020   &       -0.25   &       0.10    &       (25)&0.4551     &       0.0493  \\
V1130~Tau       &       B       &       0.1436  &       0.0025  &       0.2509  &       0.0027  &       0.740   &       0.020   &               &               &       &       &               \\
V785~Cep        &       A       &       0.0426  &       0.0028  &       0.1535  &       0.0058  &       0.340   &       0.030   &       -0.06   &       0.06    &       (26)&1.6414     &       0.2150  \\
V785~Cep        &       B       &       0.0338  &       0.0028  &       0.1377  &       0.0060  &       0.300   &       0.030   &               &               &       &       &               \\
VZ~Hya  &       A       &       0.1041  &       0.0021  &       0.1186  &       0.0017  &       0.480   &       0.040   &       -0.20   &       0.12    &       (27)&0.2798     &       0.1483  \\
VZ~Hya  &       B       &       0.0592  &       0.0027  &       0.0461  &       0.0027  &       0.240   &       0.040   &               &               &       &       &               \\
WZ~Oph  &       A       &       0.0888  &       0.0025  &       0.1464  &       0.0037  &       0.410   &       0.030   &       -0.27   &       0.07    &       (28)&1.0450     &       0.1542  \\
WZ~Oph  &       B       &       0.0864  &       0.0021  &       0.1520  &       0.0037  &       0.400   &       0.030   &               &               &       &       &               \\
Sun     &               &       0       &       4.4E-05 &       0       &       8.7E-05 &       0       &       0.00062 &       0       &       0.005   & see~text        & 0.9327        &       0.0621  \\
\hline          

\end{tabular}
\caption{\label{fgkstars}Eclipsing binaries + $\alpha~Cen$ + Sun: the obtained age and errors of the solution. {References: (1), (27) and (28)~\citet{2008A&A...487.1095C},~\citet{2008A&A...487.1081C};~(2)~\citet{2019A&A...632A..31G},~\citet{2019ApJ...872...85G};~(3)~\citet{2017A&A...597A.137K};~(4)~\citet{2011MNRAS.414.2479R};~(5)~\citet{2010A&A...516A..42C}; (6)~\citet{2018RNAAS...2...39M},~\citet{2010A&A...516A..42C};~(7)~\citet{2010AJ....139.2347L};~(8)~\citet{2020MNRAS.493.2329H};~(9)~\citet{2018ApJ...866...67T};~(10)~\citet{2010A&A...511A..22C},~\citet{1986AJ.....91..383P};~(11)~\citet{2019MNRAS.484..451H},~\citet{1986AJ.....91..383P};~(12)~\citet{2008AJ....135.1757L};~(13)~\citet{2009ApJ...707..671M};~(14) and (15)~\citet{2012Natur.481..475W};~(16), (18) e (20)~\citet{2019MNRAS.484..451H};~(17)~\citet{2020MNRAS.491.5980H};~(19)~\citet{2016ApJ...819...33L};~(21)~\citet{2020arXiv201205977S},~\citet{2012AJ....143....5S};~(22)~\citet{2016A&A...594A..92G},~\citet{2013A&A...557A.119S};~(23)~\citet{2009AJ....138.1622T};~(24)~\citet{1989A&A...211..346A},~\citet{2009MNRAS.400..969H};~(25)~\citet{2010A&A...510A..91C};~(26)~\citet{2009AJ....137.5086M}}}
\end{table*}
\section{Mass- and radius-luminosity relations: Impact of metallicity and age}
\subsection{Method} \label{method}
{To determine the intended empirical relations that allow us to compute the mass and the radius, we need to perform a multi-dimensional fit.
We follow the strategy presented in Paper I, where we use inverse problem techniques associated with the least square method. This method ensures a minimum deviation between the computed model and the input parameters \citep{Menke_1989}. The system of equations that represent the model can be written in matrix form, for example:
\begin{equation}\label{ini}
F^{obs}=Gd,
\end{equation}
where $F^{obs}$ is the matrix containing the logarithm masses in one case and the radius in the other, and $G$ the matrix with metallicity, age, and the logarithm of the luminosity. The matrix, $d,$ contains the computed model parameters that we obtain using inverse problem techniques.\\
As presented in paper I, we consider the empirical model in the form of:
\begin{equation}\label{empmod}
F^{obs}=\sum_{i=1}^N\sum_{j=1}^3 a_{i,j}x_{i}^j
.\end{equation}
The index, $i,$ describes the cases when luminosity is considered ($i=1$), when metallicity is considered ($i=2$), and, finally, when age is used ($i=3$). Depending on the number of physical quantities we used to compute the model N can have different values. Namely, 1) if just luminosity is used; 2) when luminosity and metallicity are used; and 3) if all elements are used. The index, $j,$ is the order for each parameter that goes up to third order.\\
Rewriting Eq. \ref{ini} by taking into account that our system is over-determined and then taking Eq.  \ref{empmod}, we obtain Eq. \ref{par}, which we need to solve in order to compute the desired model:\ \begin{equation}\label{par}
d^{est}=(G^TG)^{-1}G^T\log F^{obs}
.\end{equation}
\\
With this approach, the errors associated with each measurement and their influence in the final method can be considered using a Monte Carlo method.
We then performed 20000 fits, where in each case, we used randomly computed points using the standard deviation associated with each observation. In the end, we took the average of each parameter, $a_{i,j},$ and their respective standard deviation. The latter is considered the error of each parameter in the fitted model.\\
}

\subsection{Results}

\begin{figure}
\centering
\includegraphics[width=\hsize]{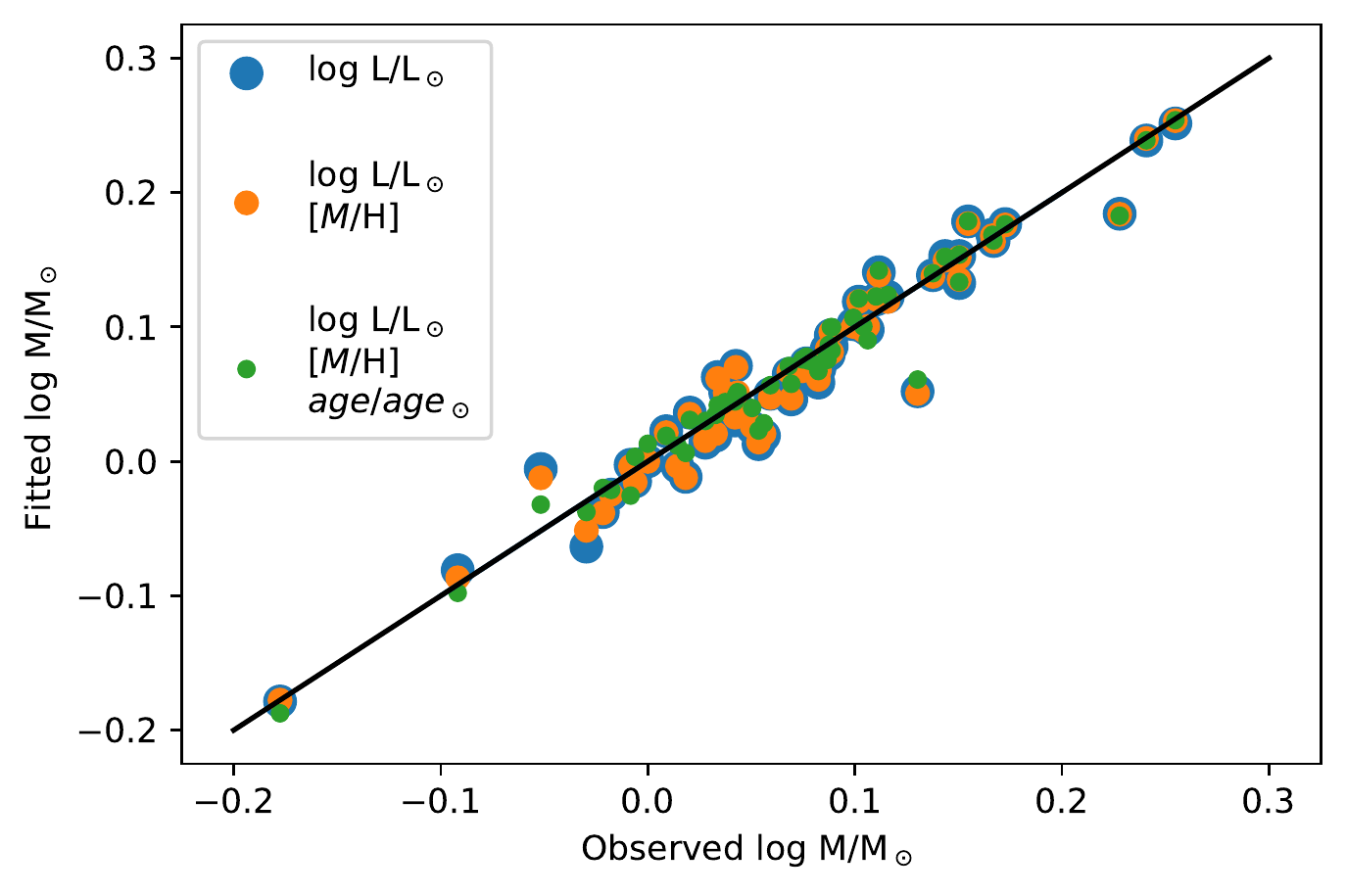}
  \caption{Fitted $\log M/M_\odot$ in function of the observed $\log M/M_\odot$. Mass-luminosity relation (blue), mass-luminosity-metallicity relation (orange), and mass-luminosity-metallicity-age relation (green).}
     \label{mass_all}
\end{figure}

In solving the Eq. \ref{par}, for the values presented in  Table \ref{fgkstars}, we can then compute the model that better fits the observed mass and radius, for the considered stars, using the different combinations for the input parameters described in Sect. \ref{method}. In plot \ref{mass_all}, we can see the results obtained estimating the mass of the stars for several N's.

Tables \ref{lum},\ref{lummet}, and \ref{lummetage} contain the obtained parameters for N=1, 2, and 3, respectively, for the empirical mass model. The correlation Pearson factor for each distribution is, respectively, 0.9663, 0.9689, and 0.9787.
In plot \ref{radi_all}, we can see the results obtained estimating the radius of the stars for several Ns.
The tables \ref{lumradius},\ref{lummetradius}, and \ref{lummetageradius} contain the obtained parameters for N=1, 2, and 3, respectively, for the empirical mass model. The Pearson correlation factor for each distribution is, respectively, 0.9825, 0.9826, and 0.9869.

\begin{table}
\small
\caption{Parameters obtained used luminosity, to fit the radius using equation \ref{empmod}.  $R^2=0.9825$ and $\sigma=0.0241$.}   \label{lumradius}
\begin{tabular}{cc}
 \hline
Parameter&values\\
\hline
$a_{1,1}$& 0.3060 $\pm$ 0.0074\\
$a_{1,2}$& 0.1036 $\pm$ 0.0067\\
$a_{1,3}$& -0.0386 $\pm$ 0.0092\\

 \hline
\end{tabular}
\end{table}

 \begin{table}
 \small
 \caption{Parameters obtained used luminosity and metallicity  to fit the radius using equation \ref{empmod}.  $R^2=0.9826$ and $\sigma=0.0240$.} 
 \label{lummetradius}
 \begin{tabular}{cc}
 \hline
Parameter&values\\
\hline
$a_{1,1}$& 0.3033 $\pm$ 0.0090\\
$a_{1,2}$& 0.1008 $\pm$ 0.0148\\
$a_{1,3}$& -0.0359 $\pm$ 0.0138\\
$a_{2,1}$& -0.0176 $\pm$ 0.0358\\
$a_{2,2}$& 0.0644 $\pm$ 0.1014\\
$a_{2,3}$& 0.2550 $\pm$ 0.4598\\
 \hline
 \end{tabular}
 \end{table}

 \begin{table}
 \small
 \caption{Parameters obtained used luminosity, metallicity and age to fit the radius using equation \ref{empmod}.  $R^2=0.9869$ and $\sigma=0.0214$.}  \label{lummetageradius}
 \begin{tabular}{cc}
 \hline
Parameter&values\\
\hline
$a_{1,1}$& 0.2908 $\pm$ 0.0133\\
$a_{1,2}$& 0.0896 $\pm$ 0.0255\\
$a_{1,3}$& -0.0170 $\pm$ 0.0195\\
$a_{2,1}$& -0.0112 $\pm$ 0.0364\\
$a_{2,2}$& -0.0603 $\pm$ 0.1123\\
$a_{2,3}$& 0.1221 $\pm$ 0.4477\\
$a_{3,1}$& 0.0084 $\pm$ 0.0196\\
$a_{3,2}$& 0.0145 $\pm$ 0.0249\\
$a_{3,3}$& -0.0057 $\pm$ 0.0074\\
 \hline
 \end{tabular}
 \end{table}

\begin{figure}
\centering
\includegraphics[width=\hsize]{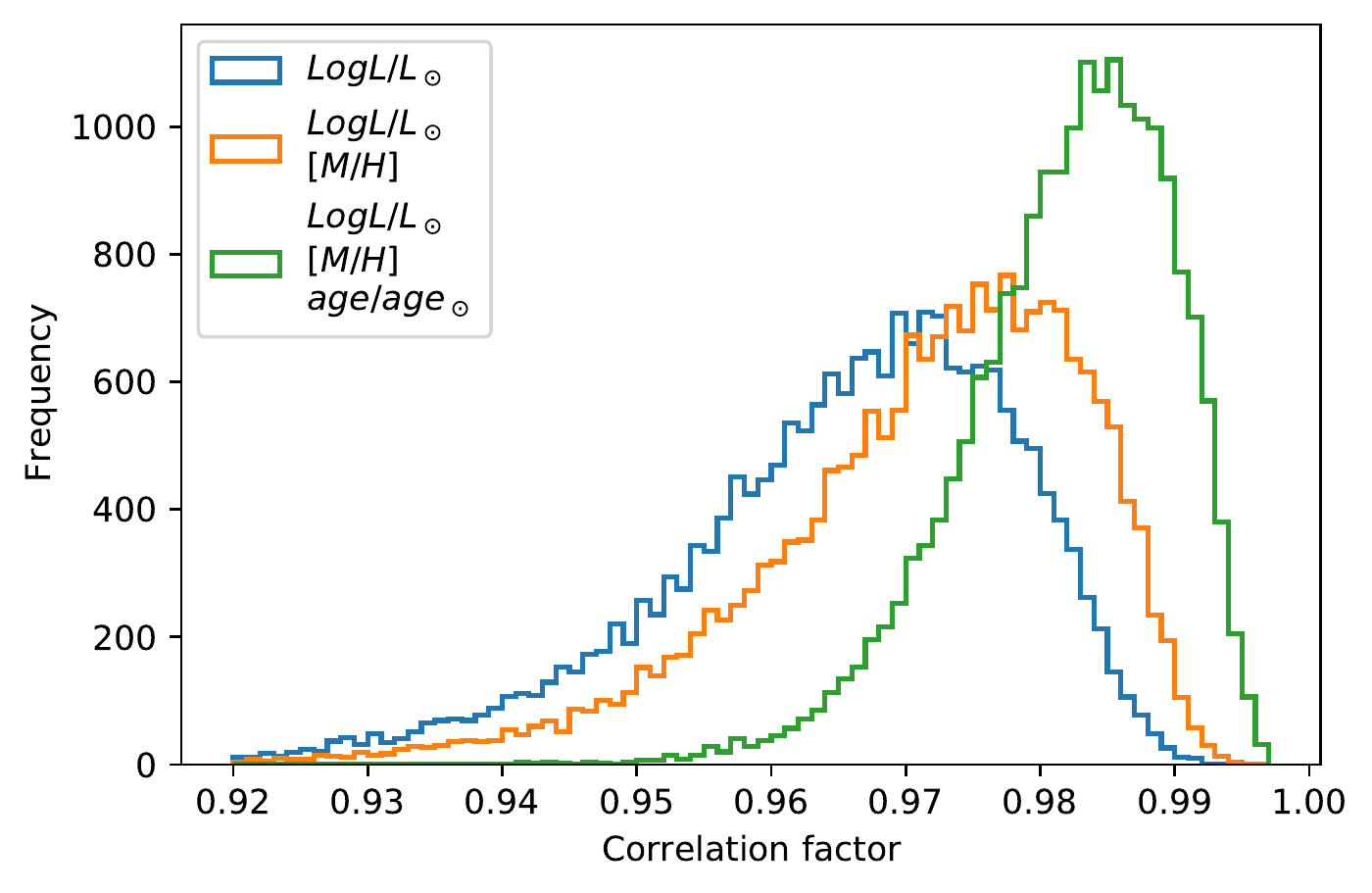}
  \caption{Distribution of the bootstrap correlation factor obtained by fitting $\log M/M_\odot$. Mass-luminosity relation (blue), mass-luminosity-metallicity relation (orange), and mass-luminosity-metallicity-age relation (green).}
     \label{mass_bootst}
\end{figure}

\begin{figure}
\centering
\includegraphics[width=\hsize]{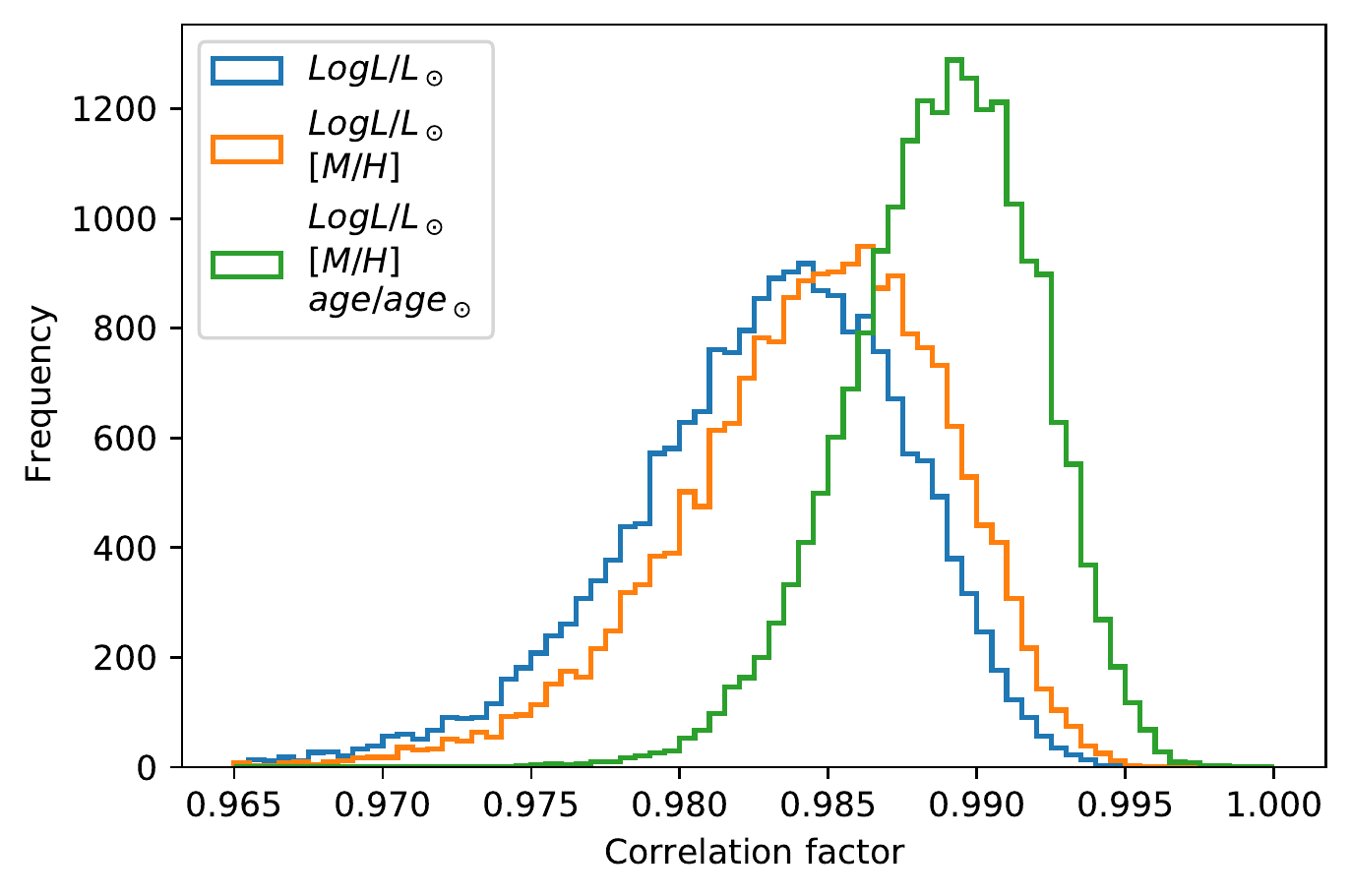}
  \caption{Distribution of the bootstrap correlation factor obtained by fitting $\log R/R_\odot$. Radius-luminosity relation (blue), radius-luminosity-metallicity relation (orange), and radius-luminosity-metallicity-age relation (green).}
     \label{radius_bootst}
\end{figure}

 To verify whether the improvement in the correlation factors obtained from the different fits is not caused solely by statistical fluctuations, we use a bootstrap \citep{efron1979}  to evaluate the significance of their differences. We re-run the fitting procedure explained before, using random stars from  Table \ref{fgkstars}, accepting repetition, until we have achieved the total number of stars. We repeated this process 20000 times, recording for each one, the obtained correlation factors.
In this case, we did not take into account the observational errors. That ensures that the input observational values used on the bootstrap are always the same.
The results of the bootstrap both for mass and radius fittings, assuming the three different combinations of input parameters, are represented in the histograms in Figs. \ref{mass_bootst} and \ref{radius_bootst}.
Analysing the results of the bootstrap, we can see that the correlation factor distribution when we use age is significantly different from the other two, showing better results. For the first two cases, the difference is not so clear. These results follow the initial estimation differences presented when a Monte Carlo method was used and is an indication of the significance of using the age in the fitting for the mass and radius of the starts.
Another aspect that we can see from the analyses of the histograms is the fact that the distributions obtained for the radius are less dispersed and they all show, on average, a better correlation than when the mass is estimated.

\begin{figure}
\centering
\includegraphics[width=\hsize]{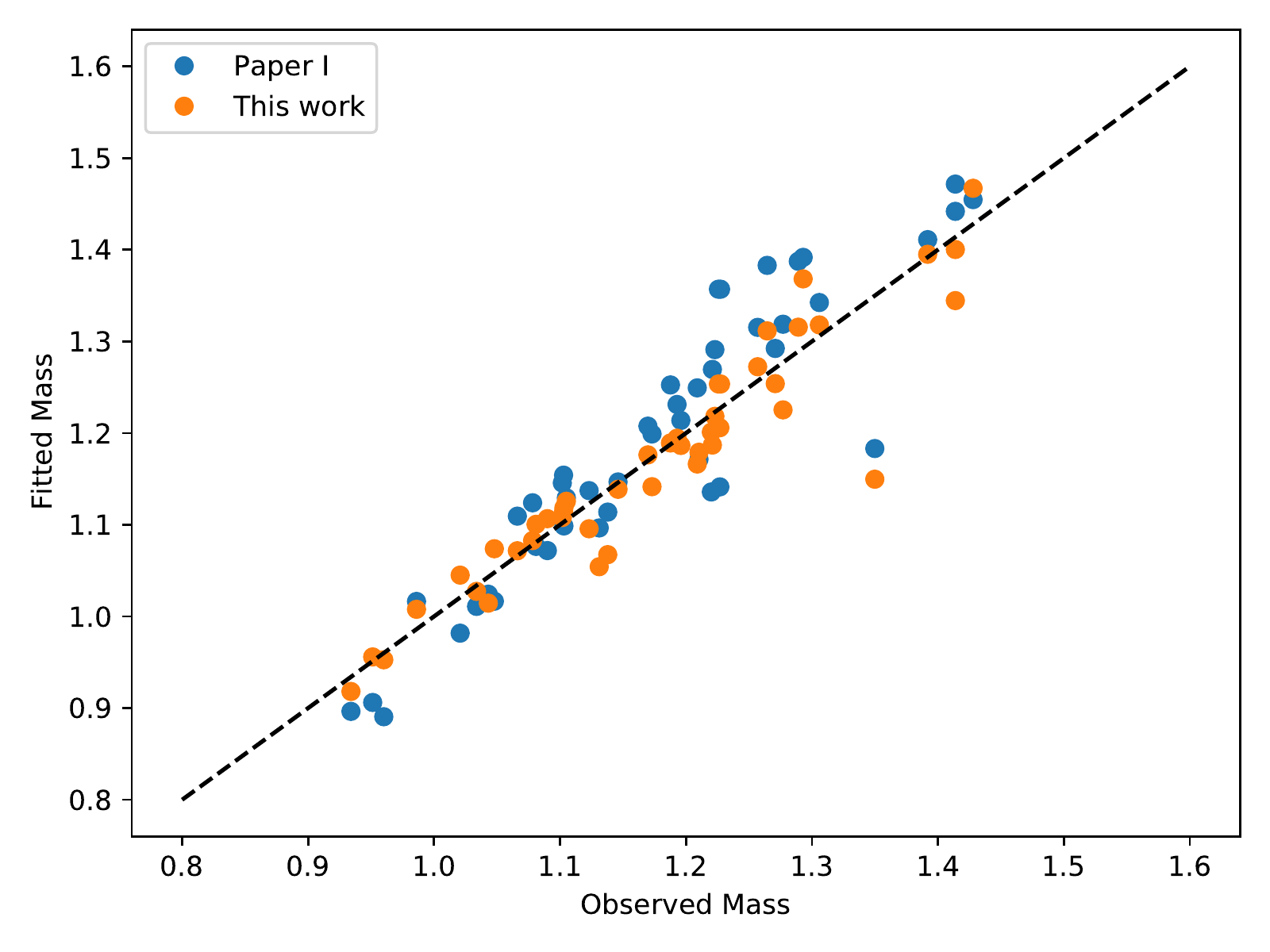}
  \caption{Mass derived by empirical relations: paper I in blue and this paper in orange, in comparison to the observed values. The black line is the one-to-one relation.}
     \label{paperI_thispaper}
\end{figure}

Below, we compare the mass- and radius-luminosity-metallicity-age relation derived in this paper (Eq. \ref{empmod} with $i=3$) with the relation derived in Paper I (for the mass)\footnote{We chose to rewrite the relation as in the Paper I there is a misprint in the last age term:  
$$\log M = 0.219(\pm 0.023)\log L/L_\odot+0.063(\pm 0.060)(\log L/L_\odot)^2-$$
$$-0.119(\pm 0.112)(\log L/L_\odot)^3+0.079(\pm 0.031)\rm{[Fe/H]}-$$
$$-0.122(\pm 0.119)[Fe/H]^2
-0.145(\pm 0.234)\rm{[Fe/H]}^3+$$
$$+0.144(\pm 0.062)Stellar Age/Age_\odot-0.224(\pm 0.104)(Stellar Age/Age_\odot)^2+$$
$$+0.076(\pm 0.045)(Stellar Age/Age_\odot)^3$$}:
we apply both relations to the observations listed in the Table \ref{fgkstars}, excluding the stars with observational mass outside the range $[0.82,1.45]M_\odot$, (Paper I relation validity). In Fig. \ref{paperI_thispaper}, we plot the comparison between two predictions. The relation derived in this paper seems to give slightly better results: the mean of the average differences between observations and predictions is $0.049M_\odot$ for Paper I and $0.028M_\odot$ in this paper.  
 
\begin{figure}
\centering
\includegraphics[width=\hsize]{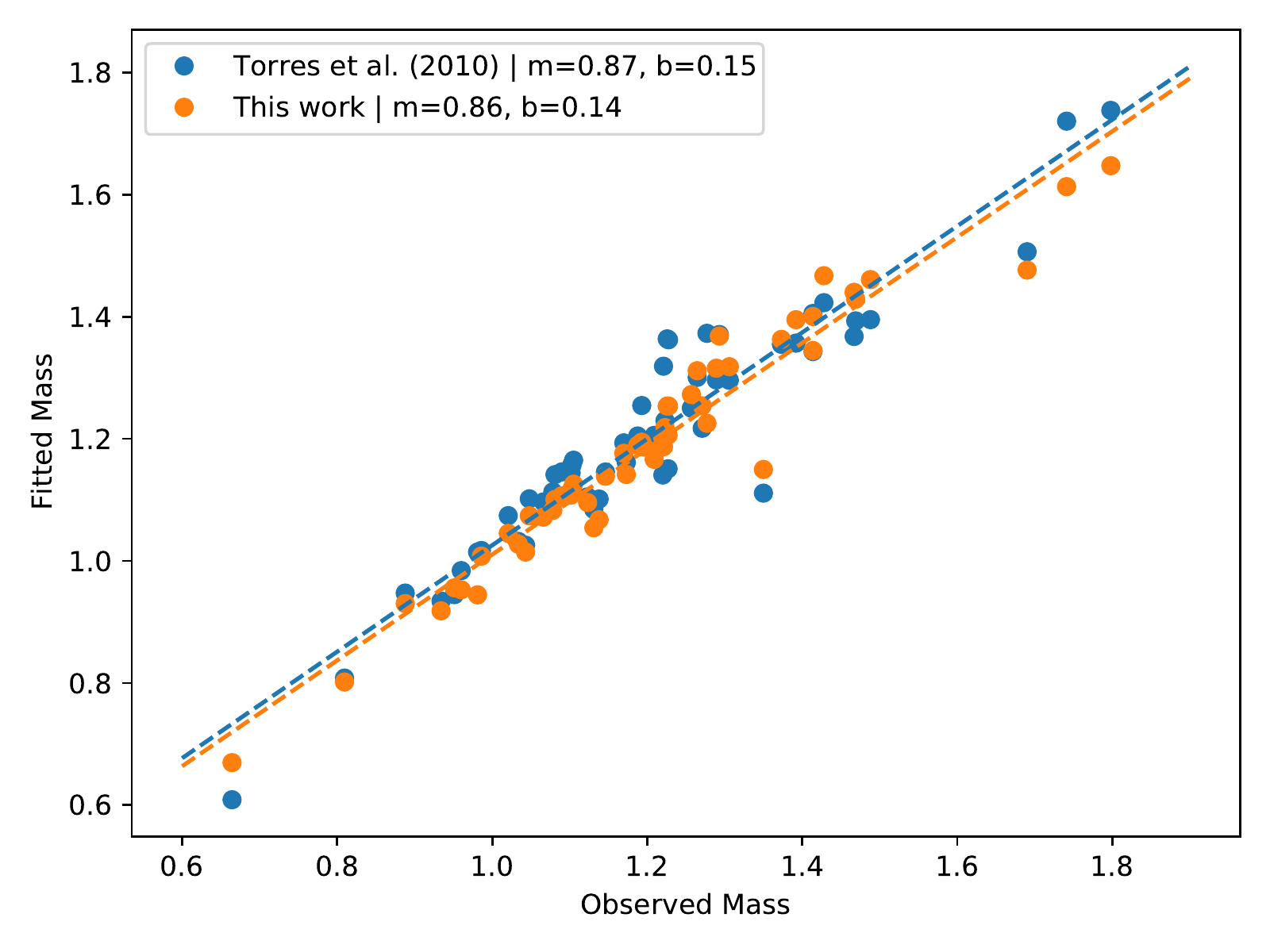}

  \caption{Mass derived by empirical relations: \citet{torres} in blue and ours in orange.}
     \label{mass_torres_ours}
\end{figure}

\begin{figure}
\centering
\includegraphics[width=\hsize]{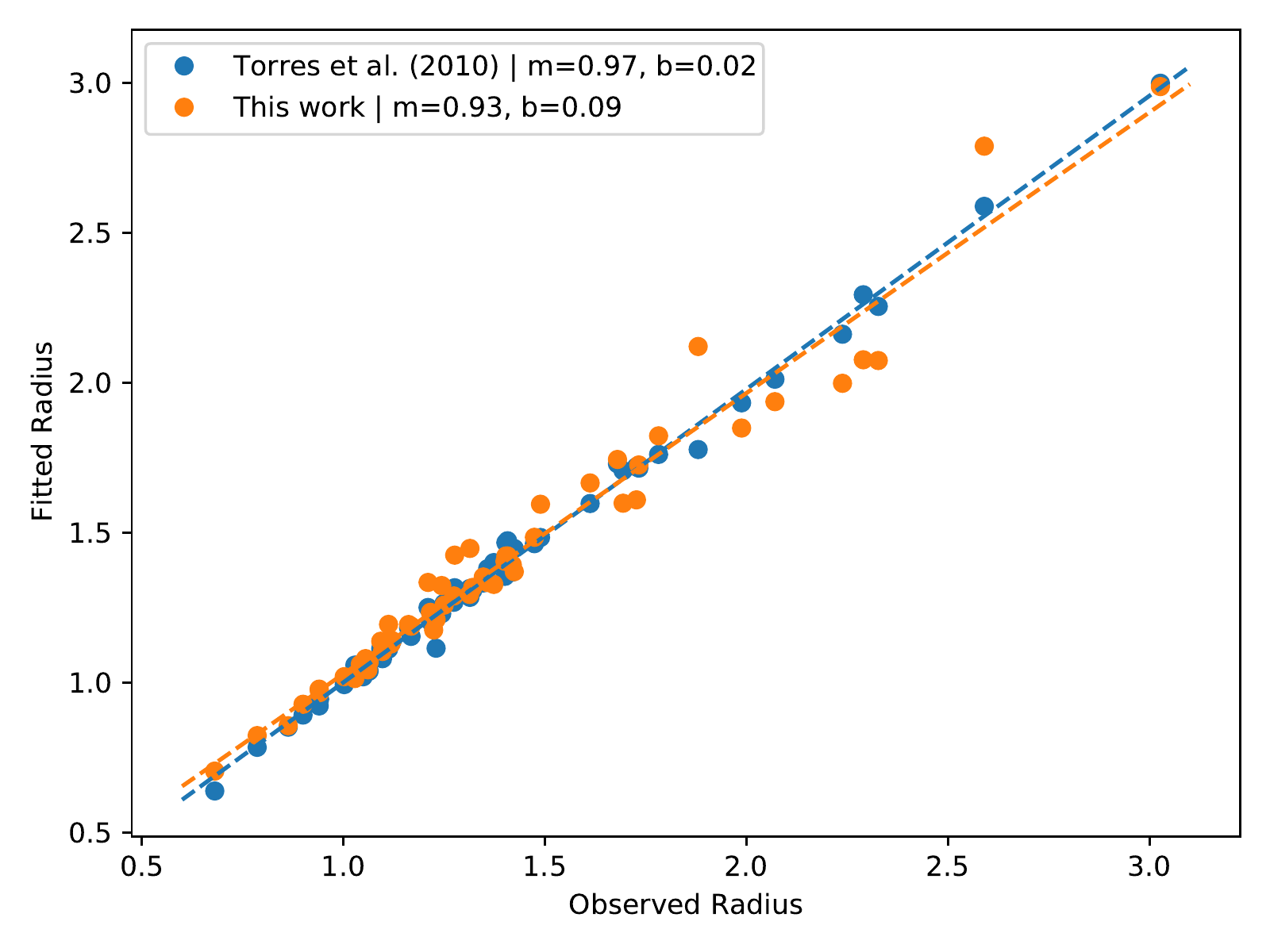}

  \caption{Radius derived by empirical relations: \citet{torres} in blue and ours in orange.}
     \label{radius_torres_ours}
\end{figure}

On the other hand, we compare our empirical relation with the one establish by \cite{torres}, both for mass and radius, using the data from the Table \ref{fgkstars} stars:
 $$ \log~M = \log~M~(T_{\rm eff},\rm{[Fe/H]},\log~g),$$
 $$ \log~R = \log~R~(T_{\rm eff},\rm{[Fe/H]},\log~g).$$
In Figs. \ref{mass_torres_ours} and \ref{radius_torres_ours}, we plot the predictions of mass and radius of our selected stars using the relations from \cite{torres} and ours in comparison with the observational data. The corresponding linear regression is also plotted. On the other hand, the mean of the absolute differences between predictions and observations is $0.048M_\odot$ and $0.026R_\odot$ for \cite{torres} and $0.035M_\odot$ and $0.059R_\odot$ for our own. It is interesting to note that these two relations are based in different approaches: \cite{torres} use only fundamental stellar observations, as $T_{\rm eff}$, $\log~g$ and [Fe/H]; on the other hand, our relation includes stellar model contributions (by the age determination). Despite that, the predictions are globally similar. 
 
Within the framework of the utility of our relations for stars with the age coming from different stellar evolutionary codes, we recompute the masses and radius using Eq. \ref{empmod} ($i=3$) and using the age determination from the source paper instead of our own. The mean of the absolute differences between predictions and observations has slight differences: $0.039M_\odot$ and $0.060R_\odot$. This means that based on an error of $\sim 1.3-1.5~Gyr$ (see Fig. \ref{graficoage}), the impact of the heterogeneity of the age determination is marginal.
 
 \subsection{Discussion}
The analysis of these results point us to two main conclusions. Firstly, the correlation factors, both for mass and radius, increases when metallicity and age are included in the respective luminosity relations. Secondly, the above increase is stronger in the mass than in the radius. The first conclusion is in agreement in our previous study, reported in Paper I. However we must emphasise that in Paper I, the correlation factor increases more dramatically from $R^2=0.830$ to $0.955$ when, respectively, only the luminosity or luminosity-metallicity-age is considered. Let us recall the metallicity range is quite different between these two papers, namely: $-0.60<$[Fe/H]$<0.40$ in Paper I against \textbf{$-0.34<$[Fe/H]$<0.27$} in this paper. The second conclusion is based on the fact that the correlation factors for the radius relations are higher than the corresponding for mass relations. In some way, the radius diagram is less dispersed than that of the mass. This can also be seen in Figures \ref{mass_torres_ours} and \ref{radius_torres_ours}. {We decide to investigate the causes, other than the obvious one related to the connection between luminosity and radius through the Stefan–Boltzmann law}. This highest dispersion in mass, rather than in radius, could  also be a consequence of an eventual larger observational error on mass than the radius. However, this is not the case here. 

\begin{figure}
\centering
\includegraphics[width=\hsize]{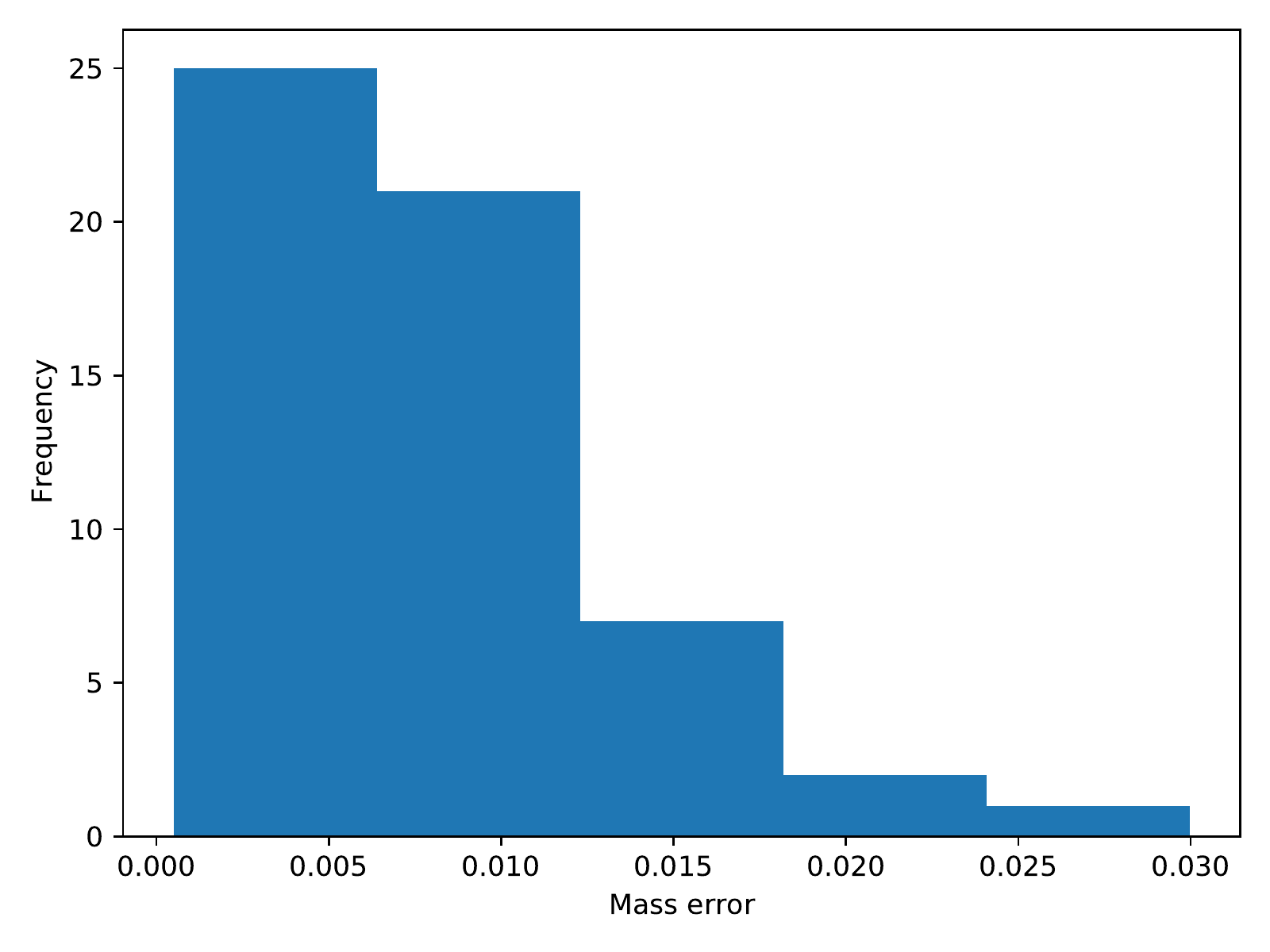}

  \caption{Histogram of observational mass errors}
     \label{mass_error}
\end{figure}

\begin{figure}
\centering
\includegraphics[width=\hsize]{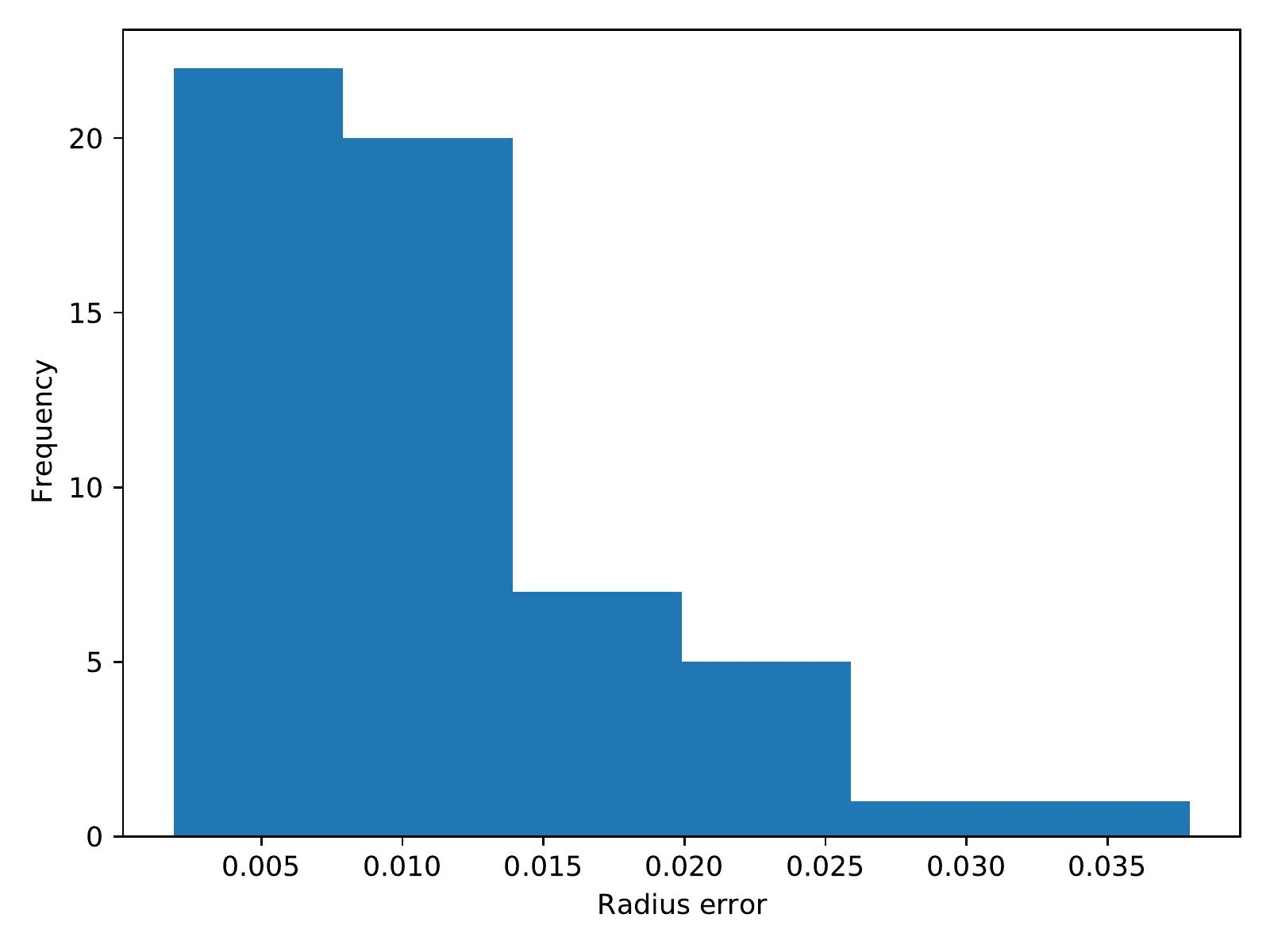}

  \caption{Histogram of observational radius errors}
     \label{radius_error}
\end{figure}

In Figs. \ref{mass_error} and \ref{radius_error}, we present the corresponding histograms for observational relative errors for mass and radius taken from {\it{DEBCat}}. The mean value is, respectively, 0.66\% and 0.88\%. By using, thus, a single sample t-test, we reject the null hypothesis of a mean mass error higher than mean radius errors (confidence level of 0.95 and p-value=0.9971). This test is based on the normal distribution for mass errors (p-value = 0.00006354, Kolmogorov-Smirnov test). Thus, observational errors do not explain the higher dispersion in the mass than as compared to the radius. Curiously, this tendency has been observed in several other studies; see Fig. 15 in the above cited paper (\cite{2018ApJS..237...21M}. In \cite{torres}, we can observe similar results, involving a scatter from these calibrations is $\sigma_{\log M} = 0.027$ and $\sigma_{\log R}= 0.014$ (6.4\% and 3.2\%, respectively) for main-sequence and evolved stars that is 'above $0.6M_\odot$'. However, also in this situation, the mean observational error on those stars are 1.2\% and 1.3\% in mass and radius, respectively. The same tendency can be observed for seismic scaling relation; see Figs. 3 and 4 from \cite{2019MNRAS.486.4612B} for more. Cross-checking these results with our above conclusions regarding the impact of age in the mass- and radius-luminosity-metallicity-age relations, it is reasonable to support the conclusion that age has a more significant contribution to mass than to the radius and this can have an impact on the highest dispersion in mass {empirical model} relations.

\section{Conclusions}
In this paper, we revisit the mass and radius-luminosity-metallicity-age relations for FGK stars. Using a multi-dimensional fit, we derive those relations for a group of 56 stars members of eclipsing binaries, including the best-known visual binary, $\alpha~Centauri$. From the $DEBCat$ catalogue, the mass and radius of those stars are known, at best, as $3\%$ to $4\%$. The luminosity for both components and metallicity (typical for Population I) for the system is also known. The age of each binary is determined by means of theoretical stellar models. The derived relations reproduce the observed values within a mean accuracy of $3.5\%$ for the mass and $5.9\%$ for the radius. We compare our results to ones published in our first paper \citet{2012Ap&SS.341..405G} and with \citet{torres} and find a similar degree of quality of predictions. Our main conclusions are: (i) the inclusion of the age improves the quality of the fit and, consequently, the predictions. This result is more clear with regard to mass than on radius; (ii) the mass-luminosity-metallicity-age relation is more dispersed than the one corresponding to the radius, as noted in several other studies (that aim to fit the observed mass or radius with such {empirical} relations). We propose that the explicit inclusion of the age tends to reduce the dispersion and, thus, to improve the quality of these relations to predict observed values. With regard to the utility of mass- and radius-luminosity-metallicity-age relations, we propose to apply it to a coeval and homogeneous (in terms of chemical composition) group of stars, such as as open clusters (e.g. in the determination of the IMF). We plan to carry out such a study in a forthcoming paper.

\begin{acknowledgements}
   We would like to thank the referee for the valuable comments which helped to improve the manuscript. The authors are grateful to the PARAM team and to J. Southworth for having made publicly available, respectively, the web interface for the Bayesian estimation of stellar parameters and DEBCat. CITEUC is funded by National Funds through FCT – Foundation for Science and Technology project
      \emph{UID/Multi/00611/2019}. J. Fernandes acknowledges funding from the POCH and Portuguese FCT –Foundation for Science and Technology \emph{ref: SFRH/BSAB/143060/2018} and visiting facilities at Niels Bohr
Institute (University of Copenhagen). This research has made use of NASA's Astrophysics Data System Bibliographic Services and the open archive "Astrophysics - arXiv".
\end{acknowledgements}

%
%

\bibliographystyle{aa}
\bibliography{BibAA}

\end{document}